# Coupling of whispering-gallery modes in size-mismatched microdisk photonic molecules

**Svetlana V. Boriskina**

*School of Radiophysics, V. Karazin Kharkov National University, Kharkov 61077, Ukraine*

Mechanisms of whispering-gallery (WG) modes coupling in microdisk photonic molecules (PMs) with slight and significant size mismatch are numerically investigated. The results reveal two different scenarios of modes interaction depending on the degree of this mismatch and offer new insight into how PM parameters can be tuned to control and modify WG-modes wavelengths and Q-factors. From a practical point of view, these findings offer a way to fabricate PM microlaser structures that exhibit low thresholds and directional emission, and at the same time are more tolerant to fabrication errors than previously explored coupled-cavity structures composed of identical microresonators. 

During the last decade, photonic molecules[1] (clusters of electromagnetically-coupled optical microcavities) have gone a long way from a useful illustration of parallels between behavior of photons and electrons and now hold promise of new insights into physics of light-matter interactions and also of a variety of practical applications, including microlasers, tunable filters and switches, coupled-cavity waveguides, sensors, etc[2-10]. The simplest PM composed of two identical optical microcavities has been widely used as a test-bed to demonstrate shift and splitting of cavity modes and formation of a spectrum of bonding and anti-bonding PM supermodes[1-4]. I have recently shown how arranging identical WG-mode microdisks into pre-designed high-symmetry configurations yields quasi-single-mode PMs with dramatically increased Q-factors[6], enhanced sensitivity to the environmental changes[7], and/or directional emission patterns[8]. In all these structures, size uniformity of microcavities is an important issue in successful realization of PM-based optical components.

The motivation for studying interactions of optical modes in a photonic molecule with size mismatch[9,10] stems from two sources. First, precise and repeatable fabrication of identical microcavities, which in many cases are tiny structures having just several microns in diameter, is highly challenging. Second, a systematic study of double-cavity PMs with various degrees of cavities size mismatch can reveal new mechanisms of manipulating their optical properties thus paving way to improving or adding new functionalities to PM-based photonic devices. Such study has never been performed before, and is a focus of this letter. Despite its simplicity, the double-cavity structure provides useful insight into the general mechanisms of WG-modes coupling and offers new design ideas for more complex structures. The Muller boundary integral equations framework previously developed by the author to model PMs composed of identical cavities[7] has been modified to study size-mismatched PMs. In the following, the term "microcavity mode" encompasses a complex value of the mode eigenfrequency and the corresponding eigenvector (i.e., modal spatial field distribution).

The PM under study is composed of a pair of side-coupled microdisks of radii $R_A$, $R_B$ and refractive indices $n_A$, $n_B$ separated by an airgap of width $w$ (Fig. 1a). The microdisks of thicknesses much smaller than their diameters are considered. Thus, instead of the 3-D boundary value problem for the Maxwell equations, we solve its 2-D equivalent. In the following analysis, we search for the TE (transverse-electric) WG-modes, which are dominant in thin disks. At wavelength $\lambda = 1.521$ μm, a 2-D cavity with radius 1.1 μm and effective refractive index $n_{eff}^{\mathrm{TE}} = 2.63$ (2-D equivalent of a 200 nm-thick GaInAsP disk)[2] supports $WG_{8,1}$-mode with one radial field variation and eight azimuthal field variations (Q = 5243). This mode (like all other WG-modes in circular cavities) is double-degenerate due to the symmetry of the structure.

WG-mode degeneracy is removed if two (or more) cavities are brought close together[1-9], and four non-degenerate WG-supermodes of different symmetry appear in the double-disk PM spectrum instead of every WG-mode of an isolated cavity. Fig. 2 (b and c) shows the wavelength migration and Q-factors change of these modes with the change of the radius of one of the cavities. The modes are labeled according to the symmetry of their field patterns along the *y*- and *x*- axes, respectively (e.g., OE supermode has odd symmetry with respect to *y*-axis and even symmetry with respect to *x*-axis). OE and OO modes are termed "anti-bonding" modes, while EO and EE modes are termed "bonding" ones. Bonding and anti-bonding supermodes group into nearly-degenerate doublets as seen in Fig. 1b. The values of real parts of eigenfrequencies of two modes in a doublet are so close to each other that they cannot be distinguished (Fig. 1b), while their imaginary parts differ, resulting in different Q-factors of these supermodes (Fig. 1c). Thus, in practice only two peaks are observed in a symmetrical double-cavity PM lasing spectrum (see Fig. 2 in Ref. 9), where the narrow high-intensity peak corresponds to the high-Q anti-bonding mode doublet, and the wider low-intensity peak corresponds to the bonding one.

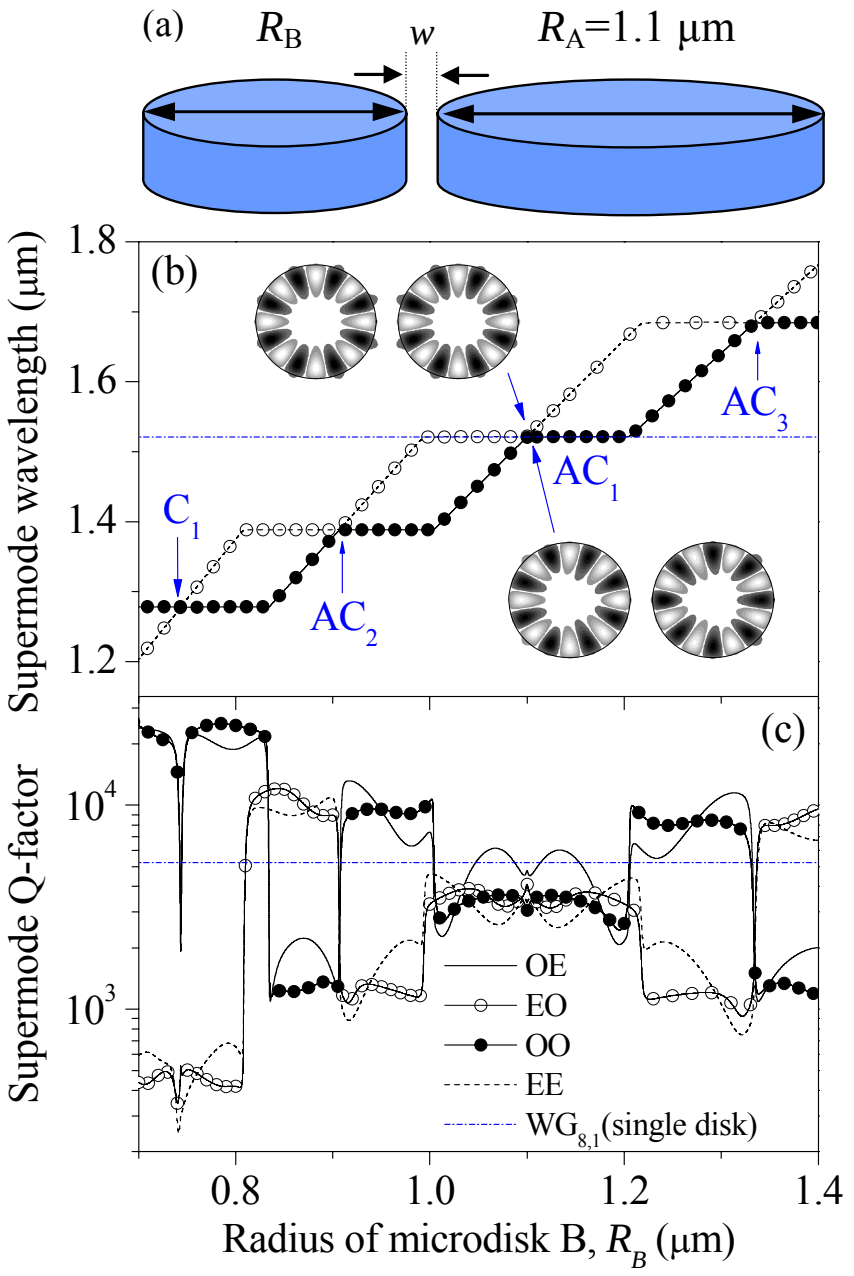

Fig 1. (a) A geometry of a PM composed of two microdisks of different radii; (b) wavelengths migration and (c) Q-factors change of PM supermodes as a function of the radius of disk B ($R_A$ = 1.1 µm, $w$ = 400 nm). The insets show the magnetic field distribution of the bonding (EE) and the anti-bonding (OE) $WG_{8,1}$ supermodes in the symmetrical ($R_A = R_B$ = 1.1 µm) PM. Here and thereafter, corresponding characteristics of the $WG_{8,1}$ mode of an isolated cavity with radius 1.1 µm are plotted for comparison (dash-dot line).

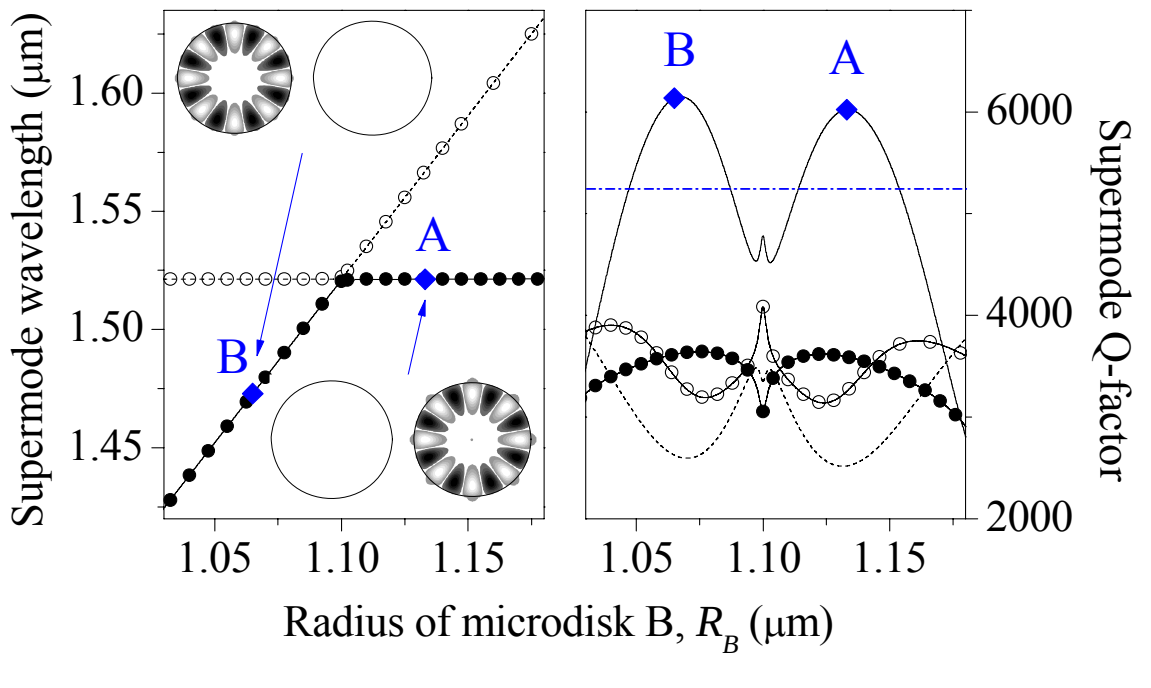

Fig. 2. Supermodes wavelengths (left) and Q-factors (right) in the vicinity of anti-crossing point AC$_1$ ($R_B = R_A$). Mode switching (see the modal near-field distributions at points **A** and **B** shown in the insets) at the anti-crossing point and Q-factor enhancement of one of the supermodes can be observed.

Careful study of Figs. 1 b,c reveals a number of so-called exceptional points (corresponding to certain values of the varied parameter), where PM supermodes couple following either crossing (C) or avoided crossing (AC) scenarios. The behavior of wavelengths and Q-factors of the four supermodes in the vicinity of these exceptional points is shown in more detail in Figs. 2-4 (for the points AC$_1$, AC$_2$, and C$_1$, respectively). The phenomenon of coupling of complex eigenvalues of matrices dependent on parameters under the change of these parameters is of a general nature and is observed in many physical systems[11]. Usually, frequency anti-crossing (crossing) is accompanied by crossing (anti-crossing) of the corresponding widths of the resonance states. Furthermore, at the points of avoided frequency crossing (points AC$_1$-AC$_3$ in Fig. 1), eigenmodes interchange their identities, i.e., Q-factors and field distributions.

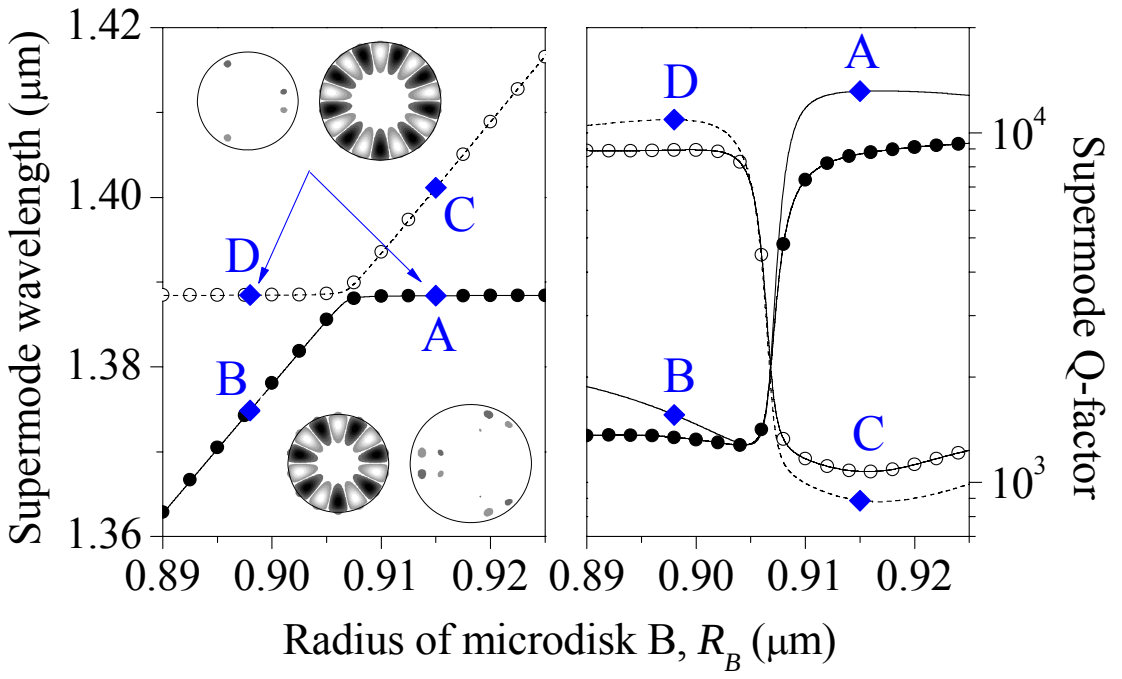

Fig. 3. Supermodes wavelengths (left) and Q-factors (right) in the vicinity of anti-crossing point AC$_2$. Wavelengths repulsion accompanied by the linewidths crossing is observed. The insets demonstrate mode switching at the anti-crossing point (the modal near-field distributions shown in the upper(lower) insets correspond to the complex frequency values at the points labeled as **A** and **D** (**B** and **C**), respectively.

In the context of coupling of WG-modes in microcavities, this interchange offers exciting new prospects for manipulating the PM optical characteristics, e.g. for realization of optical flip-flops[9]. For example, the coupling of modes with avoided frequency crossing scenario observed in Figs. 2 and 3 makes possible switching of field intensity between two microdisks. To realize such switching in practice, carrier-induced refractive index change of one of the disks induced by nonuniform pumping can be used. This effect was observed experimentally[11] in a PM composed of nearly-identical microdisks (similar to the case shown in Fig. 2). If the microcavities are severely size-mismatched, their WG-modes couple with the frequency crossing scenario. This situation is demonstrated in Fig. 4, and the numerical data indicate that such coupling may spoil significantly the Q-factors of the high-Q modes in the larger microdisk. However, optical coupling between cavities with strongly detuned WG-modes makes possible broad spectral transmission effects in coupled resonator optical waveguides (CROWs)[10], coupled-resonator induced transparency[12], and significant reducing of CROW bend radiation loses[13].

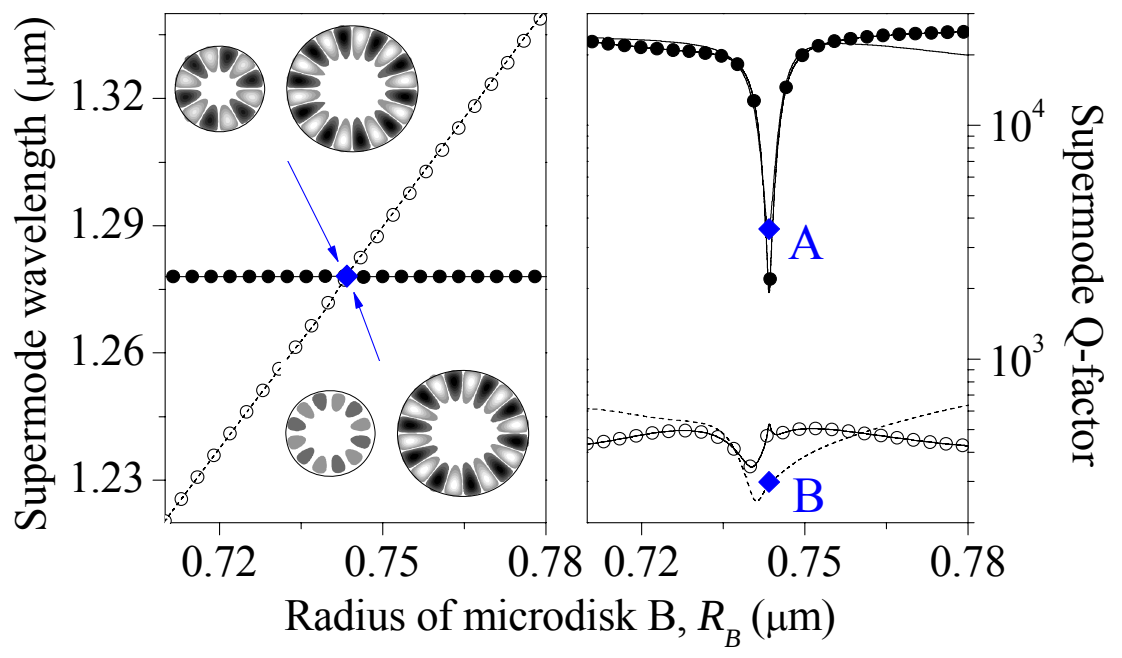

Fig. 4. Supermodes wavelengths (left) and Q-factors (right) in the vicinity of the crossing point $C_1$. Wavelength crossing accompanied by damping of the high-Q supermodes is observed. The insets show supermodes near-field portraits at the crossing point.

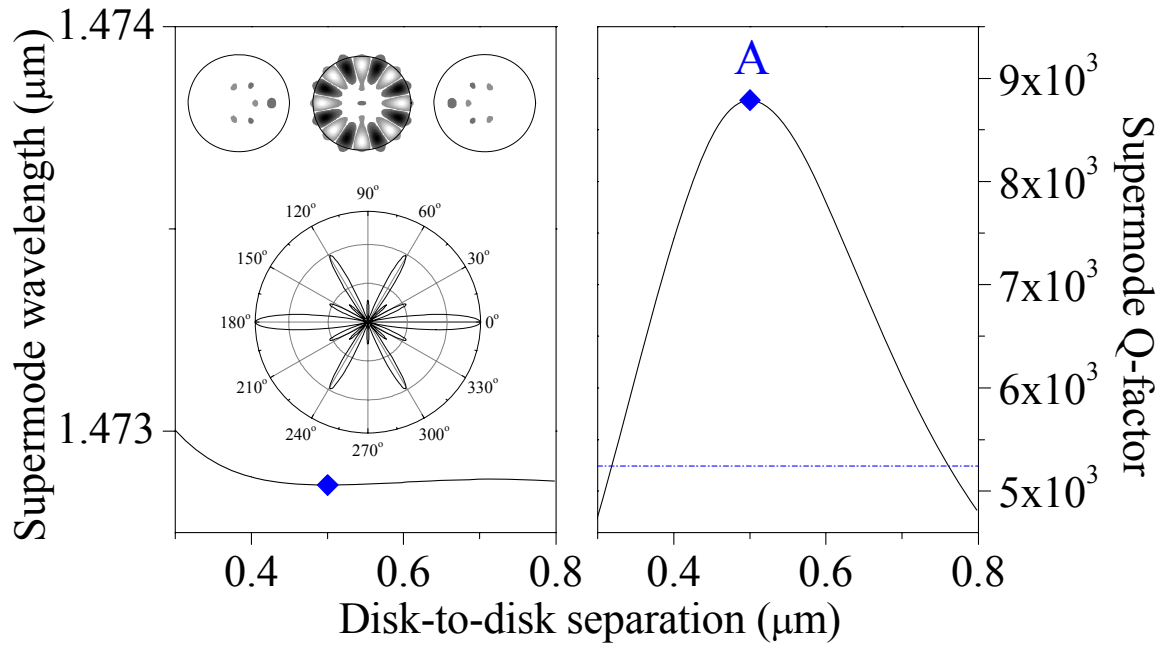

Fig. 5. Resonance wavelength (left) and Q-factor (right) of an anti-bonding WG-supermode in a three-disk PM. The central disk of radius 1.065 μm is separated from the side disks of radii 1.1 μm by airgaps of 400 nm width. Supermode near-field portrait and directional far-field emission pattern at the point labeled as **A** are shown in the inset.

Finally, enhancement of the Q-factor of a single supermode in a double-disk PM in comparison to its single-cavity value can be observed in Fig. 2 in a relatively wide range of cavities radii detuning: 14 nm < $\Delta R$ < 53 nm ($\Delta R = |R_A - R_B|$). Note that all the other PM supermodes have significantly lower Q-factors in this range of the parameter change. This effect offers a way for selective enhancement of the Q-factor of a single supermode that (unlike symmetry-enhanced Q-factor boost in polygonal PMs)[6,7] does not rely on exact cavity size-matching. A possible realization of a PM-based structure designed by making use of this mechanism of selective mode enhancement is presented in Fig. 5. It consists of three coupled microcavities, with the central cavity radius detuned by $\Delta R$ = 35 nm from the side cavities radii. By adjusting the width of the airgaps between microcavities, noticeable Q-factor enhancement of one anti-bonding supermode is achieved without shifting the supermode wavelength (Fig. 5). Furthermore, such PM demonstrates directional light emission, which cannot be achieved in isolated WG-mode microdisks (see inset to Fig. 5). Our studies also indicate that this directional emission pattern is preserved if the disk-to-disk distance is varied.

It should also be noted that other system parameters can be tuned to manipulate resonance wavelengths and Q-factors of microcavities through mode coupling at exceptional points. Among these are: the refractive index of the cavity substrate, and the size and/or position of a hole pierced in the cavity, which can be adjusted to enhance a WG-mode Q- factor[14,15] or to achieve directional emission on a high-Q WG-mode[16].

In summary, a comprehensive numerical study was performed to elucidate the mechanisms of modes coupling in PMs with various degrees of cavities size mismatch. The study offers an alternative approach to design novel PM-based components with improved functionalities. In contrast to PM structures composed of identical cavities that may require fabrication accuracy beyond the capabilities of modern technology, the proposed approach does not rely on precise cavities size-matching to achieve the desired device performance. This approach paves the way for new designs of more complex PM structures and arrays, which may eventually lead to new capabilities and applications in micro- and nano-photonics.

I wish to thank Vasily Astratov for discussions and Jan Wiersig for bringing his recent paper[16] to my attention. This work has been partially supported by the NATO Collaborative Linkage Grant CBP.NUKR.CLG 982430. Svetlana Boriskina's e-mail address is SBoriskina@gmail.com.